\documentclass[a4paper,10pt,3p,twocolumn]{elsarticle}

\usepackage{subcaption}
\usepackage{graphicx}
\usepackage{hyperref}
\usepackage{siunitx}
\usepackage[version=4]{mhchem}
\usepackage{caption}
\usepackage{subcaption}
\journal{Astroparticle physics journal}
\newcommand{\Xmax}{\ensuremath{X_\mathrm{max}\;}}

\begin{document}

\begin{frontmatter}

\title{A Deep Learning-based Reconstruction of Cosmic Ray-induced Air Showers}

\author{M.~Erdmann}
\ead{erdmann@physik.rwth-aachen.de}
\author{J.~Glombitza}
\author{D.~Walz}
\address{Physics Institute 3A, RWTH Aachen University, D-52056 Aachen, Germany}

\begin{abstract}
We describe a method of reconstructing air showers induced by cosmic rays using deep learning techniques. 
We simulate an observatory consisting of ground-based particle detectors with fixed locations on a regular grid.
The detector's responses to traversing shower particles are signal amplitudes as a function of time, which provide information on transverse and longitudinal shower properties.
In order to take advantage of convolutional network techniques specialized in local pattern recognition, we convert all information to the image-like grid of the detectors.
In this way, multiple features, such as arrival times of the first particles and optimized characterizations of time traces, are processed by the network.
The reconstruction quality of the cosmic ray arrival direction turns out to be competitive with an analytic reconstruction algorithm.
The reconstructed shower direction, energy and shower depth show the expected improvement in resolution for higher cosmic ray energy.
\end{abstract}


\end{frontmatter}

\section{Introduction}

The successes of the deep learning ansatz in handwriting recognition~\citep{Hinton,Ciresan}, speech recognition~\citep{Yu}, and competitions with humans regarding image identification and playing games~\citep{ILSVRC2012,ResNet1,Go} motivates its application in various areas of fundamental research.
For an overview of deep learning techniques see~\citep{review}.

In physics research, tasks like reconstructing particle kinematics or signal separation from background processes often require algorithms that take multiple observables into account in order to obtain a few parameters of interest. 
Machine learning techniques are frequently used for such complex problems, with boosted decision trees and shallow neural networks being among the most popular methods.

Concepts of deep neural networks have recently been investigated for challenges in astroparticle and particle physics. 
Various network designs have been successfully applied in simulation studies, e.g., to reconstruct the neutrino flavor in neutrino nucleus interactions~\citep{NoVA}, to extract a new exotic particle or Higgs boson signal from a background dominated data sample~\citep{exotics,higgs,kaggle}, to identify the underlying parton flavor of a jet or measure jet substructure~\citep{jetflavor,Baldi:2016fql}, or to assign jets to the underlying partons in top quark-associated Higgs boson production~\citep{ttH}.

In this work we investigate deep learning methods for reconstructing cosmic ray properties from simulated air showers.
As our observatory of the air showers we use ground-based particle detectors located on a hexagonal grid, each delivering charge measurements as a function of time.

Reconstruction of the cosmic ray properties from air showers includes arrival direction and energy.
In addition, the atmospheric depth of the shower maximum is of interest to obtain information on the cosmic ray mass.
All this information is commonly extracted from the data using algorithms based on physics arguments which were developed by astroparticle physicists.

As an alternative method, we exploit deep learning techniques to reconstruct the above-mentioned properties of cosmic rays from simulated data.
Our aim is to investigate the abilities of the network to learn about various aspects encoded in the data, and to evaluate the reconstruction quality of the cosmic ray arrival direction, energy, and shower maximum.

We designed the network to study the following aspects separately.
One aspect is the spatial distribution of signals on the ground.
Further information is contained in the time of the first signals arriving at different detector stations on the ground.
A third aspect is the time trace of the detector signals which encodes arrival times of different particles from different phases of the shower development.

This work is structured as follows.
First we introduce a parametrized simulation of air showers.
We then explain the details of the network architecture and discuss how the data are input to the network.
We evaluate the quality of air shower reconstructions obtained with the network using various aspects of the data before presenting our conclusions.

\section{Parametrized simulation of air showers}

For our investigations we developed a parametrized air shower simulation code, and a corresponding detector simulation inspired by~\citep{corsika,parameterized}.
The simulation code is published in \citep{gitlab}.
This program enables the efficient generation of a large number of simulated events for network training and evaluation on one hand, and direct control of the information contained in the data on the other hand.
On output the simulation delivers the same information as obtained by a ground-based particle detector such that there are no limitations in extending our studies to detailed air shower simulations.

The results of the air shower simulations are parametrized directly in terms of the signal distributions obtained in the detectors. 
The detectors are placed on a hexagonal grid with a spacing of \SI{1500}{m} and are located at a height of \SI{1400}{m} above sea level, motivated by the Pierre Auger Observatory~\citep{Auger}.

We consider only the electromagnetic part and the muon part of the air shower, both of which depend on the arrival directions, energies, and nuclear masses of the cosmic rays. 

The energy $E$ of the cosmic rays is randomly selected between 3 and \SI{100}{EeV} following a power law $E^{-1}$. 
The mass composition $A$ of the cosmic rays consists of \ce{H}, \ce{He}, \ce{N}, \ce{Fe} with equal fractions. 
The arrival directions $(\theta,\phi)$ are chosen following an isotropic distribution with zenith angles of up to $\theta=\SI{60}{\degree}$. 
Finally, the impact point of the shower on the surface, the shower core, is randomly sampled around the central detector.

\begin{figure}[h!]
\centering
\includegraphics[width=0.45\textwidth]{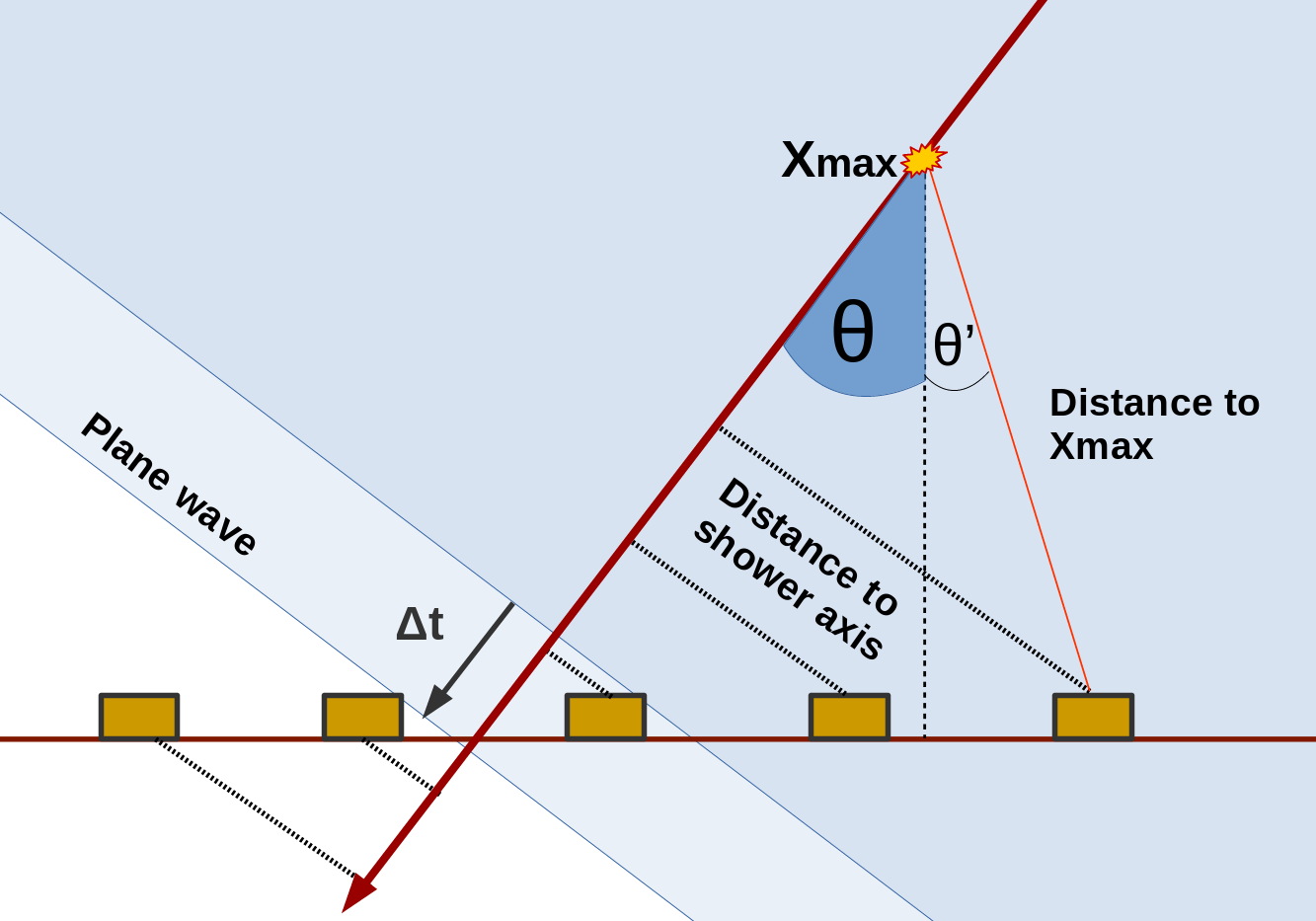}
\caption{Parametrized air shower simulation with zenith angle $\theta$ and reference point \Xmax
to determine the detector distances to the shower axis and to estimate atmospheric absorption effects.
A planar shower front is used to determine the arrival times of the first particles at the detectors.
\label{fig:airshower}}
\end{figure}

For the chosen values of the energy and mass $(E,A)$, two decisions are taken which influence the detector signals.
The first decision is the spatial reference point for calculating distances of the detectors to the shower, see Fig.~\ref{fig:airshower}. 
The distribution of the maximum of the atmospheric shower depth \Xmax can be well approximated by a Gumbel distribution $G(E,A)$ \citep{Gumbel}. 
The randomly selected \Xmax value from $G(E,A)$, together with the arrival direction $(\theta,\phi)$ and the shower core, defines the reference point for all subsequent geometry calculations. 
Starting here, the movement of the shower front is approximated in terms of a plane wave.

The second decision to depend on the chosen $(E,A)$ values is the relative energy distribution between the electromagnetic and the muonic components of the shower. 
For proton primaries, we set the electromagnetic energy to $70\%$ of the cosmic ray energy and the muon part to $30\%$, respectively. 
For all other nuclei, the relative energy contained in the muonic component is up-scaled by a factor $A^{0.15}$~\citep{Matthews}. 
The signal distributions in the detectors of both the electromagnetic and the muonic components are calculated separately and are finally superimposed. 

Each detector is supposed to have its own clock recording a universal time $t_{0}$ 
when the trigger starts the electronics to record particle signals as a function of time $t$.
The time $t_{0}$ is calculated according to the movement of the shower front.

The time trace $S(t)$ of particle signals arriving at the detector 
contains 80 intervals of \SI{25}{ns} size. The shape of the signal distributions is approximated by a log-normal distribution for both the electromagnetic and the muon signal ($j=\mu,em$), following \citep{parameterized}:
\begin{equation}
F_{j}(t) = \frac{1}{\sqrt{2\pi\sigma^2}\,x_j} \; \exp{\left( -\frac{( \ln(x_j) - \tau_j )^2}{2\sigma^2}\right)}
\label{eq:timetrace}
\end{equation}
with $x_j = (t - \Delta t_j) / t'$.
Here $t'$ is a reference time to cancel time units, $\Delta t_j$ a time offset specified below, $\tau_j$ the location parameter of the log-normal distribution and $\sigma=0.7$ the shape parameter.
To simulate the effect of a lateral distribution function of the shower, the location parameter
\begin{equation}
\tau_{j} = \ln{\left[a_{j}+ b_{j}\;\left(\frac{r}{r_\circ}\right)^c\; \left(1-d_{j}\;\frac{\Delta X}{X_\circ} \right)
\right]}
\label{eq:reference_time}
\end{equation}
depends on the transverse distance $r$ of the detector to the shower axis.
To also include the absorption of shower particles in the atmosphere, the difference $\Delta X$ between the atmospheric depth of the detector and the above-mentioned reference point of the shower at \Xmax is calculated. 
$a_j$ contains a global offset, and $b_j$ and $d_j$ are weights for the lateral distance and the absorption effects, which are different for the muonic and the electromagnetic components. The choice of parameters has been adapted to approximate distributions obtained from full shower simulations \citep{parameterized}
($a_\mu=80$, $a_{em}=50$, $b_{\mu}=140$, $b_{em}=200$ , $c=1.4$, $d_{\mu}=0.2$, $d_{em}=0.1$, $r_\circ=\SI{750}{m}$, $X_\circ=\SI{1000}{g/cm^2}$). 
For the electromagnetic component a time offset is added, reading
\begin{equation}
\frac{\Delta t_{em}}{t'} =\frac{3}{2} \left( \exp(\tau_{em})-\exp(\tau_{\mu}) \right)
\end{equation}
with respect to the muonic component, $\Delta t_\mu / t' = 0$.

The total energy deposit in the detector, or total signal strength $S'_{0}$ respectively, respects the same effects of the lateral distribution function and the atmospheric absorption as presented in (\ref{eq:reference_time}):
\begin{equation}
S_{j,0}' = S_{j,0}\;\left(\frac{r}{r'}\right)^{\alpha_{j}}\; \left(\;\frac{\Delta X}{X'}\right)^{\beta_{j}}.
\label{eq:energy}
\end{equation}
Here the parameters are $\alpha_{\mu}=-4.7$, $\alpha_{em}=-6.1$, $\beta_{\mu}=0.1$, $\beta_{em}=0.4$, $r'=\SI{1000}{m}$, $X'=\SI{100}{g/cm^2}$.

For each detector $i$, the signal distribution $S_{i}(t)$ as a function of time is obtained by a Monte Carlo method. 
Random values are chosen according to (\ref{eq:timetrace}) where the number of drawn values is proportional to the signal strength $S'_{j,0}$ (see eq. \ref{eq:energy}). To prevent diverging signals a maximum signal is set with respect to the shower core.

Finally, effects of noise contributions are added to the signal distributions. 
In each interval of the time trace a uniformly distributed noise between $0-5\%$ of the signal and a Gaussian background noise are added to the interval ($\mu=1.2$, $\sigma=0.6$).

Also the trigger times $t_{0}$ of each detector are subjected to noise effects expressed 
in terms of Gaussian distributions where the width varies between $\SI{0.06}{\mu s}$ and $\SI{0.32}{\mu s}$ depending on the total signal $(S'_{0} = S'_{em,0} + S'_{\mu,0})$ in the detector.

\begin{figure}[h!]
\captionsetup[subfigure]{aboveskip=-1pt,belowskip=-1pt}
\begin{centering}
\begin{subfigure}[b]{0.495\textwidth}
\includegraphics[width=\textwidth]{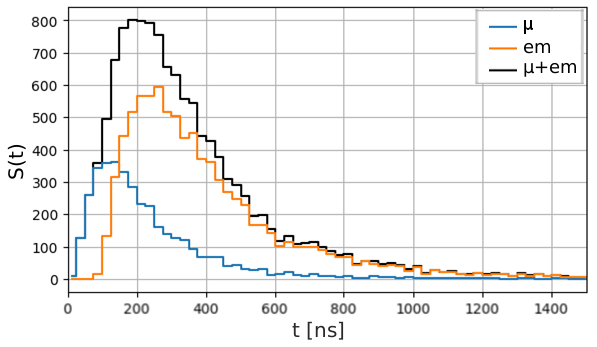}
\subcaption{}
\label{fig:trace_a}
\end{subfigure}
\begin{subfigure}[b]{0.495\textwidth}
\vspace*{0.3cm}
\includegraphics[width=\textwidth]{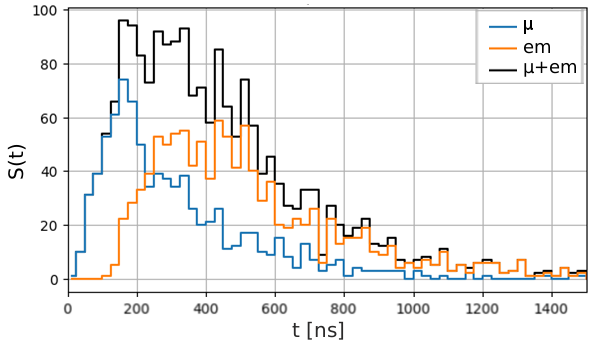}
\subcaption{}
\label{fig:trace_b}
\end{subfigure}
\caption{Example time traces $S(t)$ in arbitrary units as a function of time $t$ in units of ns, 
a) detector near the shower core, 
b) detector further away from shower core.}
\label{fig:trace}
\end{centering}
\end{figure}

Fig.~\ref{fig:trace} shows example distributions of the simulated time traces for a single event in a detector 
near the shower core (Fig.~\ref{fig:trace_a}), and in a detector at some distance from the shower core 
(Fig.~\ref{fig:trace_b}). 
While the muonic component always arrives early, the electromagnetic component is relatively late for the detector located further away from the shower core.

\section{Data preprocessing}

The available data for the reconstruction of cosmic ray-induced air showers consist of the time $t_{0,i}$ when the first shower particles arrive at each detector $i$ and the time traces $S_i(t)$ of the detectors.

In order to fully exploit advantages arising from network techniques specialized in local pattern recognition and to achieve good stability and efficiency in optimizing the network, the data are transformed to suitable formats.
These concern geometry aspects on one hand, and the numerical ranges of the data values on the other hand.

\paragraph{Geometry of the detector array}

The success of the so-called convolutional network technique~\citep{convolution} in computer vision relies on the analysis of sub-regions of an image with a spatially invariant operation.
The convolution technique can be understood as sliding a small filter, say, $3 \times 3$ pixels, in regular steps over the image and applying it to the respective image subregions.  

In a figurative sense, measuring a cosmic ray-induced air shower with an arrangement of identical particle detectors can be considered as taking a pixelized image of the shower footprint.
In contrast to a digital camera where the pixels typically form a Cartesian grid, many particle detectors are arranged on a hexagonal grid.
Hence, a suitable representation of the hexagonal detector grid is needed in order to apply the convolution technique for investigating subregions in the shower footprint.

\begin{figure}[h!]
\centering
\hspace*{-0.09cm}
\includegraphics[width=0.48\textwidth]{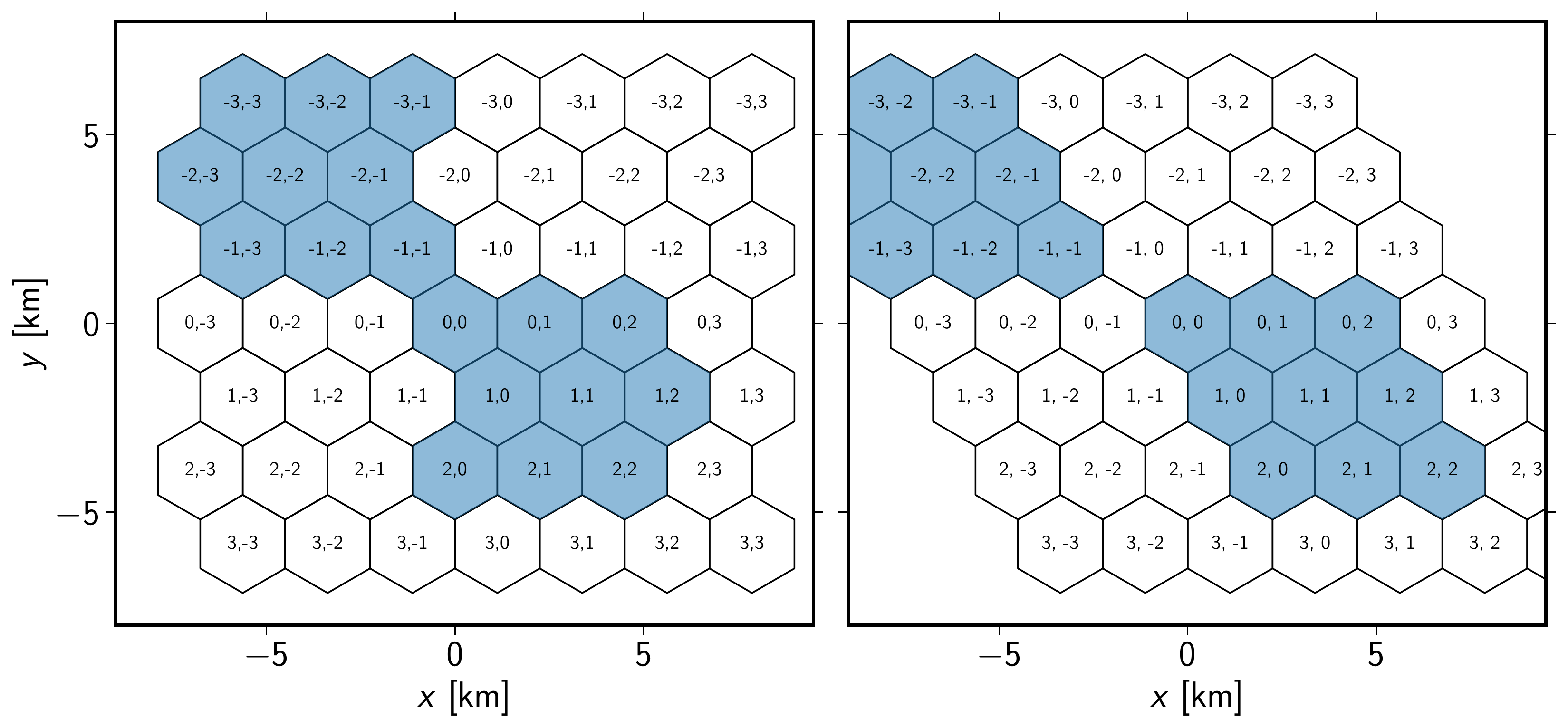}
\caption{
Comparison of representations for detectors arranged on a hexagonal grid, offset coordinates (left) and axial coordinates (right).
Hexagons outline the spatial extent of the pixel corresponding to each grid point, indexed by the given coordinates.
Shown in blue is a $3 \times 3$ filter applied at two different positions.}
\label{fig:grid}
\end{figure}

Fig.~\ref{fig:grid} shows two such representations, using offset and axial coordinates, respectively.
While offset coordinates allow for a more compact representation of rectangular sections than axial coordinates, a possible disadvantage arises from the changing filter shape depending on the position where the filter is applied.
This is illustrated on the left of Fig.~\ref{fig:grid} where $3 \times 3$ filters at two different positions are shown in blue.
When comparing the two positions, one can see that in offset coordinates the pixels in the top and bottom row of the filter are at different positions relative to those in the middle row.

Such a changing relationship may be problematic for learning spatial correlations.
However, in practice we observed no difference in reconstruction performance between offset and axial coordinates.
It is possible that translational invariance is less important in this application because the shower core is always centered close to the image center.
Consequently, we choose offset coordinates to allow for a more compact representation.
With this mapping, the triggered stations for all showers in the simulated dataset are contained within a $9 \times 9$ grid, which is the size used in the following.

\paragraph{Normalization of arrival time data}

The arrival times of the first particles in all detectors of the observatory provide a two-dimensional map containing information that can be used to extract, e.g., the shower direction.
This will be discussed in a later section.

Prior to injecting data to the network it is advantageous to limit the numerical range of the data values
by a suitable transformation.
The trigger times $t_{0,i}$ of the first particles arriving at the detectors $i$ are averaged within an event,
and are zero-centered by taking this average value $\tau_{\rm event}$ as a reference value:
\begin{equation}
\label{eq:t0}
\tilde{t}_{0,i} = \frac{t_{0,i} - \tau_{\rm event}}{\sigma_{t, {\rm data\; set}}}
\end{equation}
The spread $\sigma_{t, {\rm data\; set}}$ of these time values is obtained by the standard deviation
of the transformed time values $(t_{0,i}-\tau_{\rm event})$ using all stations and all events of
the data sample.
Stations without a signal are set to zero, $\tilde{t}_{0,i} = 0$.

\paragraph{Normalization of the amplitudes}

The lateral energy distribution of the shower approximately follows a power law~\citep{Auger}.
In order to obtain a more linear input to the network, the amplitudes $S_i(t)$ of each time trace are transformed by a logarithmic function:
\begin{equation}
\tilde{S}_i(t) = \log_{10}[S_i(t) + 1]
\end{equation}
Here an offset of 1 is used to map stations without signal to zero, $S_i(t)=0\rightarrow \tilde{S}_i(t)=0$.

Also, the total signal of the time trace is calculated for each detector by summing the amplitudes $S_{\rm tot,i}=\sum S_i(t)$ of all 80 time intervals.
The same transformation as for the single amplitudes is applied here:
\begin{equation} \label{eq:Stot}
\tilde{S}_{\rm tot,i} = \log_{10}[S_{\rm tot,i}+1]
\end{equation}
The transformed total signals deliver a two-dimensional map which contains information that can be used to extract, e.g., the total energy contained in the shower. This will also be discussed in a later section.

\paragraph{Summary of the input data}

The information available to the network on input are the 2 two-dimensional maps of the transformed total signals $\tilde{S}_{tot}$ and the transformed arrival times $\tilde{t}_{0}$ of the first particles. 
In addition, we have the time traces of the detectors $i$ with the transformed amplitudes $\tilde{S_i}(t)$ which contain information on the longitudinal shower development.

Instead of using $\tilde{S_i}(t)$ directly we will use a sub-network to extract a number of features $f_{j}$ characterizing these distributions.
For example, one feature may characterize the initial rise of the distribution, another one the length of the distribution etc.
During the training process, the network will learn features of the time traces which are optimal for predicting the requested shower property.
The exact choice of features remains hidden.
Every feature $f_{j}$ will be evaluated for all detectors such that the $f_{j}$ provide $N=10$ two-dimensional maps.
These feature maps can be combined directly with the above maps of $\tilde{S}_{tot}$ and $\tilde{t}_{0}$ in further network layers.

\section{Network architecture and training}

\begin{figure}[t!h]
\centering
\includegraphics[width=\columnwidth]{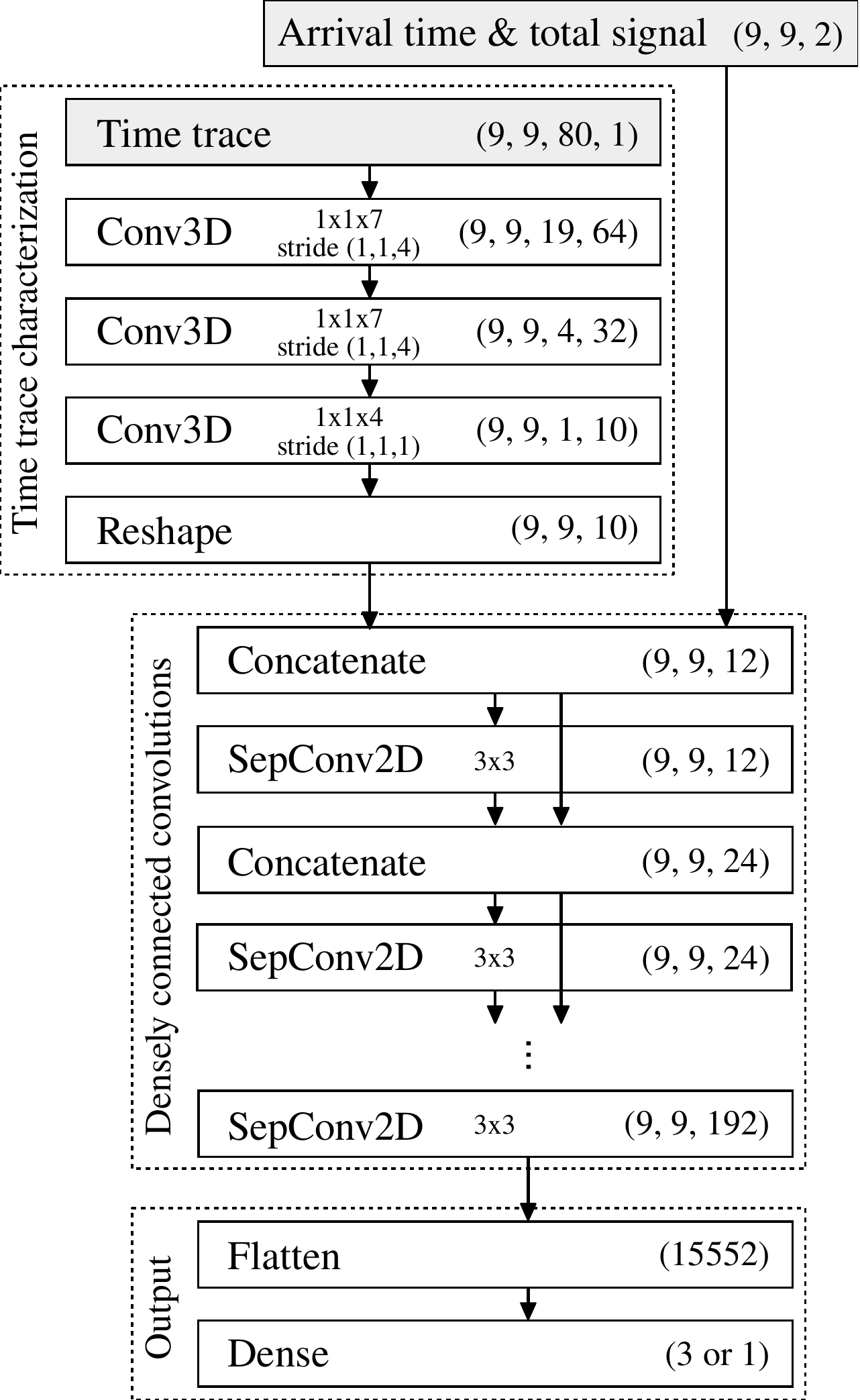}
\caption{Network architecture:
A series of 3D convolutions characterizes the time trace of each detector by 10 features which are  then stacked with the arrival time and the total signal. 
The main part consists of five separable 2D convolution layers whose in- and output are concatenated for the next layer. This creates a shortcut of the original input and the output of each convolution layer to all subsequent layers. A final dense layer outputs the shower property.
}
\label{fig:architecture}
\end{figure}

To reconstruct the individual shower properties, shower direction, energy and $X_\mathrm{max}$, we set up and train separate networks of the same structure.
The network architecture, depicted in Fig.~\ref{fig:architecture}, consists of three functional parts.
The first part characterizes the time traces at each detector station.
Then follows the main network part performing a series of convolutions on the stacked maps of arrival time, total signal and extracted time trace features.
Finally, a dense layer predicts one or three output values depending on the reconstruction task.
As deep learning framework we use Keras~\citep{Keras} with TensorFlow as backend~\citep{tensorflow}.
Our network implementation is published in \citep{gitlab}.

\paragraph{Time trace characterization}

The first network part takes as input a tensor of shape (9, 9, 80, 1) holding for each of the $9 \times 9$ detector stations the time trace over 80 time bins.
Here, the last dimension of size 1 is required for technical reasons.

In order to characterize these time traces, three consecutive layers of 1D convolutions are applied to each detector station separately.
Technically this is implemented using 3D convolutions with filters of size $1 \times 1$ in the spatial dimensions, thus performing the same operations on the time traces.

For the three layers we use \{64, 32 and 10\} filters of size \{7, 7 and 4\}, sliding over the time traces with a stride of \{4, 4 and 1\} without padding.
By striding, each convolution sees the time trace at a lower resolution and broader scale. 
After the third convolution the time dimension is consumed and the output tensor is of shape (9, 9, 1, 10) with the fourth dimension holding the 10 extracted features $f_j$.
The consumed time dimension is removed by reshaping to (9, 9, 10).

This network part can be interpreted as a single function $f(\tilde{S}(t)) = f_j$ characterizing the time trace of each detector by a small number of features.
Since the entire network is trained as a whole, the function will learn to extract those $10$ features which are most useful for the following network in the given task.

\paragraph{Densely connected convolutions}

For the main network part, the $9 \times 9$ images of extracted time trace features are concatenated with arrival time and total signal, yielding a tensor of shape (9, 9, 12).
Two design choices are central to this network part: depth-wise separable convolutions and shortcuts by densely connected layers.

Due to the different physical meaning of each of these 12 feature maps, we expect that spatial correlations within a feature map are sufficiently decoupled from correlations between features, such that it is preferable not to map them jointly.
We incorporate this assumption into the network structure by making use of depth-wise separable convolutions~\citep{Xception}.
These layers first perform spatial convolutions separately on each feature map, before correlating all feature maps pixel-wise in a second step.
This factorization of spatial and cross-feature correlations allows for increased computational and parameter efficiency compared to standard convolutions.

Shortcuts have proven to be an essential feature for successfully training deep neural networks~\citep{ResNet1,ResNet2}.
For air shower reconstruction we use a network with densely connected layers based on the DenseNet architecture~\citep{DenseNet}.
In this architecture, the output of a layer is provided as input to each subsequent layer.
Technically this is implemented by concatenating the input of a layer with its output, which then forms the input to the following layer.

The benefit is twofold:
First, these shortcuts allow the signal and gradients to better propagate in the forward and backward pass, respectively.
Second, the unaltered feature maps corresponding to arrival time, total signal and time trace features are input through these dense connections to each weight layer in the main network part as shown in Fig.~4.
This allows the network to better represent mathematical expressions from standard reconstruction methods.
In the same way, the derived features are input to all following layers, which enforces feature reuse.

Concretely, we use a series of five separable convolutions and concatenations.
The number $f$ of output features in each separable convolution is chosen equal to the input features, hence $f = \{12,\,24,\,48,\,96,\,192\}$, and the output of each concatenation is $2f$.
The filter size is $3 \times 3$ and padding is used to keep the spatial size constant.

\paragraph{Output layer}

Finally, the $(9, 9, 192)$ output tensor of the network's main part is flattened and projected with a single, fully connected weight layer without activation function onto the regression target.
For the tasks of reconstructing the energy or the depth of shower maximum, a single output predicts 
$E / \SI{}{EeV}$ and $\Xmax / (\SI{}{g/cm^2})$, respectively.
For shower direction reconstruction we avoided having the network predict $(\phi,\theta)$ directly since the spherical angles contain a pole at $\theta=0$ and a periodicity at $\phi = 0 = 2\pi$.
Instead, the network predicts the Cartesian $(x,y,z)$ unit vector.
Our tests showed that normalizing the predicted vector to unit length before the loss function does not result in an improved angular resolution.

\begin{figure*}[t!]
\captionsetup[subfigure]{aboveskip=-1pt,belowskip=-1pt}
\begin{centering}
\begin{subfigure}[b]{0.495\textwidth}
\includegraphics[trim={0 0 2cm 2.1cm},clip,,width=\textwidth]{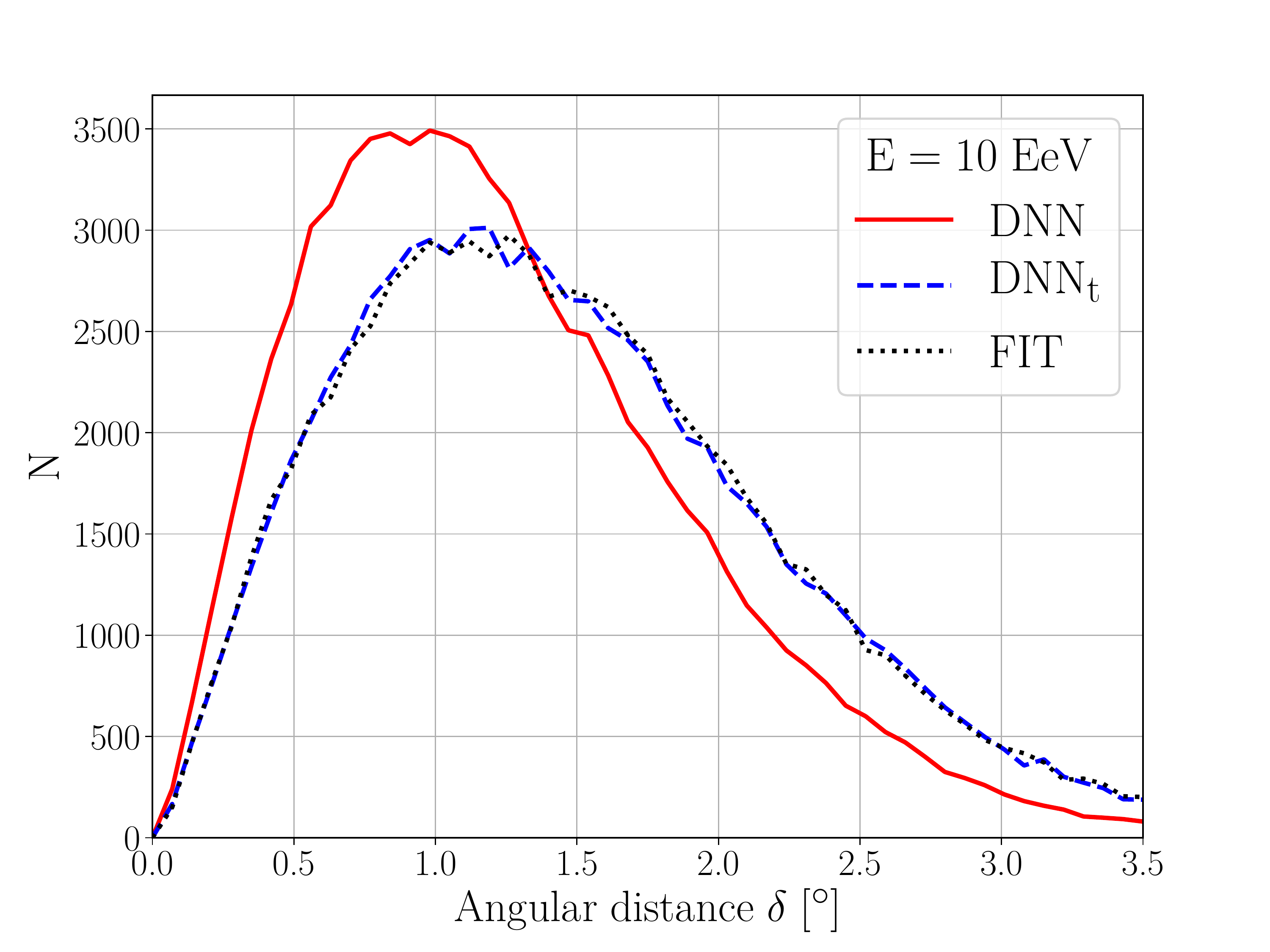}
\subcaption{}
\label{fig:angular_a}
\end{subfigure}
\hfill
\begin{subfigure}[b]{0.495\textwidth}
\includegraphics[trim={0 0 2cm 2.1cm},clip,,width=\textwidth]{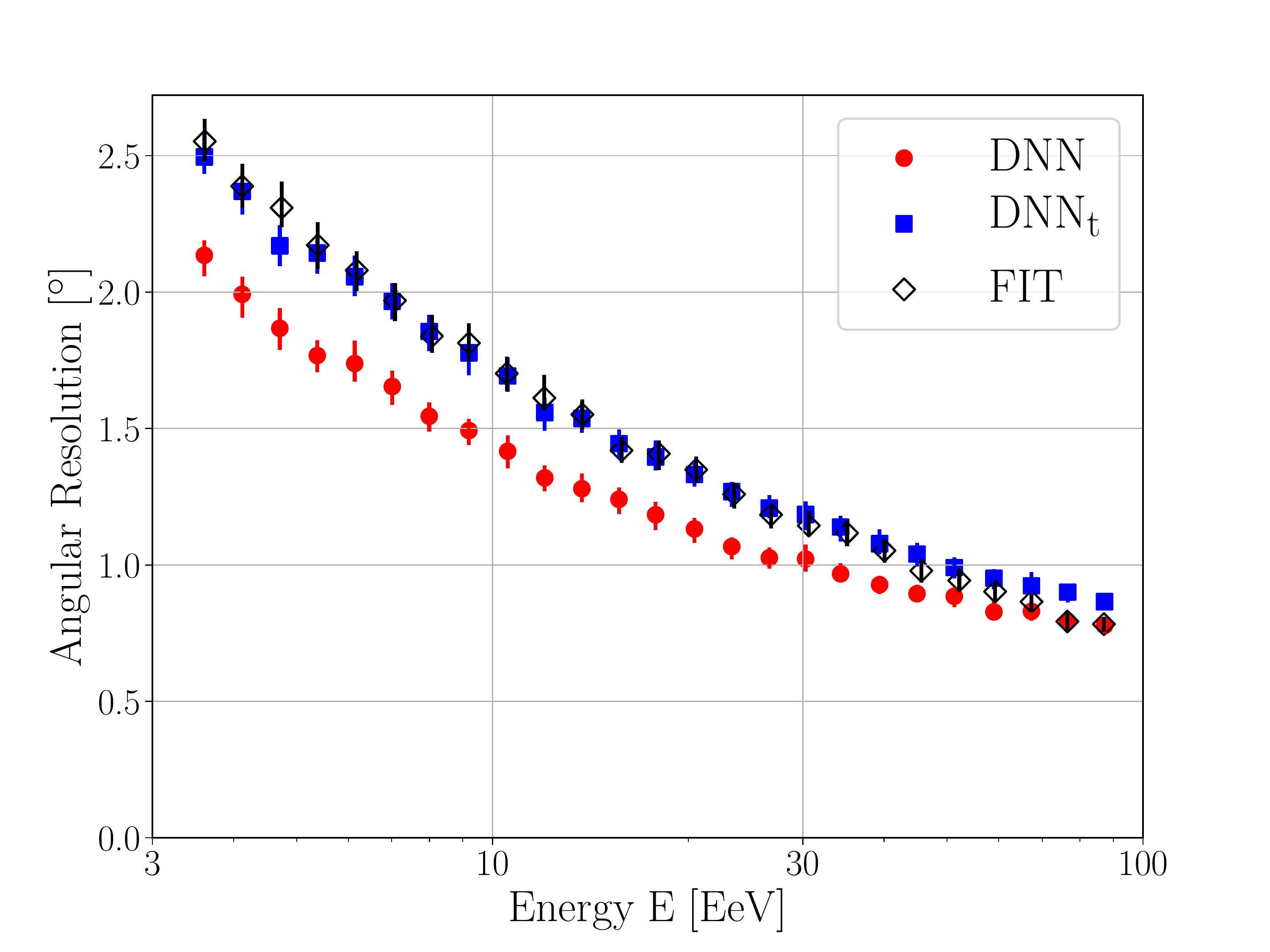}
\subcaption{}
\label{fig:angular_b}
\end{subfigure}
\caption{
(a) Angular distance between reconstructed and true shower directions for cosmic rays with \SI{10}{EeV}.
(b) Angular resolution, defined by the 68th percentile of the angular distance distribution, as a function of cosmic ray energy. 
Curves and symbols refer to the analytic fit (black diamond symbols, dotted curve), the network with time information only (blue square symbols, dashed curve) and the network with full information (red circular symbols, solid curve).
Error bars show the statistical uncertainty of the resolution.
}
\end{centering}
\end{figure*}

\paragraph{Discussion of the network design}

We use ReLU as nonlinearity after all weighted layers, except for the output layer~\citep{review}.
As alternatives we considered $\tanh$ and ELU~\citep{ELU}, however we found no improvement in terms of training efficiency or final performance.

Tested variations of the main network part include a simple series of convolutions as well as additional residual shortcuts, each of varying depth and number of parameters, and considering both standard and separable convolutions.
None of these variations showed improved results.

Interestingly, our tests also showed no benefits of using signal normalization, such as batch normalization~\citep{BatchNorm}, in any part of the network.
When reconstructing the shower direction, batch normalization even had a contrary effect.
We conjecture that reconstructing the shower direction from the arrival times on the ground requires a level of fine-tuning that is negatively affected by the noise induced through batch normalization.

With the training procedure described below, the networks show little generalization error, thus regularization is not required.
Using dropout~\citep{dropout} in combination with an increased network capacity showed no significant improvement.

\paragraph{Training}

The reconstruction task is set up as a regression problem using the mean squared error between predicted and target value as loss function.
Since ReLU activations are used, we initialize all weights according to the MSRA scheme~\citep{MSRA}.
As optimizer we use ADAM~\citep{ADAM} with standard values for the initial learning rate $\alpha = 10^{-3}$ and for the first two moments' exponential decay rates $\beta_1=0.9$, $\beta_2=0.999$.

The network has 84,547 parameters for reconstructing scalar shower properties (energy, shower maximum), and 115,653 parameters for reconstructing the arrival direction.
It is trained on 358,000 showers in batches of 132. Each training is run on a single NVIDIA GeForce GTX 1080 GPU. The training time per epoch is $\sim \SI{100}{s}$ and inference on a batch of 132 showers takes \SI{0.04}{s}.

During training, the validation loss is monitored on a fixed validation set of 2,000 showers and used for reducing the learning rate.
Specifically, the learning rate is reduced by a factor of around 0.5 whenever the validation metric reaches a plateau.
Each training is run over 120 epochs.

For each task we train five networks and select the best performing network based on the validation set.
The reconstruction performance is then evaluated on two independent test sets. One set contains 40,000 simulated showers with an $E^{-1}$ energy spectrum between 3 and \SI{100}{EeV}, the other set 80,000 simulated showers of \SI{10}{EeV} energy.
Both sets have a mixed composition of \ce{H}, \ce{He}, \ce{N}, \ce{Fe} with equal fractions.
The performance in these test sets is described in the following sections.

\section{Arrival direction}
The reconstruction of the arrival directions is based primarily on the arrival times of the first shower particles at the detectors.

As a benchmark, we perform a plane fit by using the arrival times together with the locations of the detectors.
In Fig.~\ref{fig:angular_a}, the angular distance of the fitted shower directions and the true directions are shown for 80,000 events of the mixed composition (\ce{H}, \ce{He}, \ce{N}, \ce{Fe} with equal fractions) and an energy of \SI{10}{EeV} by the dotted black curve (FIT).
The angular resolution amounts to \SI{1.45}{\degree} expressed in terms of the $68\%$ quantile of the distribution.

\begin{figure*}[t!]
\captionsetup[subfigure]{aboveskip=-1.5pt,belowskip=-1pt}
\begin{centering}
\begin{subfigure}[b]{0.495\textwidth}
\includegraphics[trim={0 0 2cm 2.1cm},clip,width=\textwidth]{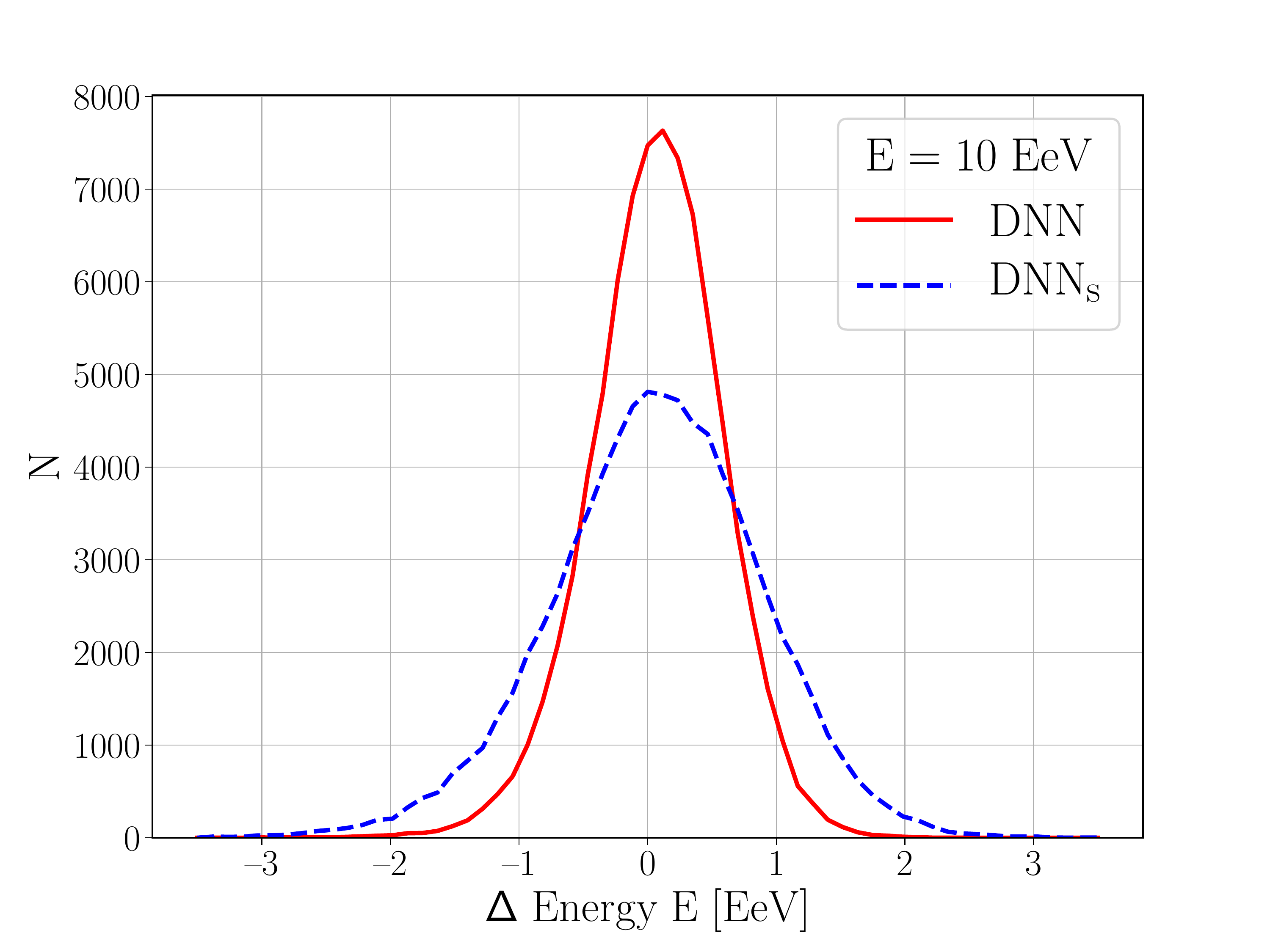}
\subcaption{}
\label{fig:energy_a}
\end{subfigure}
\hfill
\begin{subfigure}[b]{0.495\textwidth}
\includegraphics[trim={0 0 2cm 2.1cm},clip,width=\textwidth]{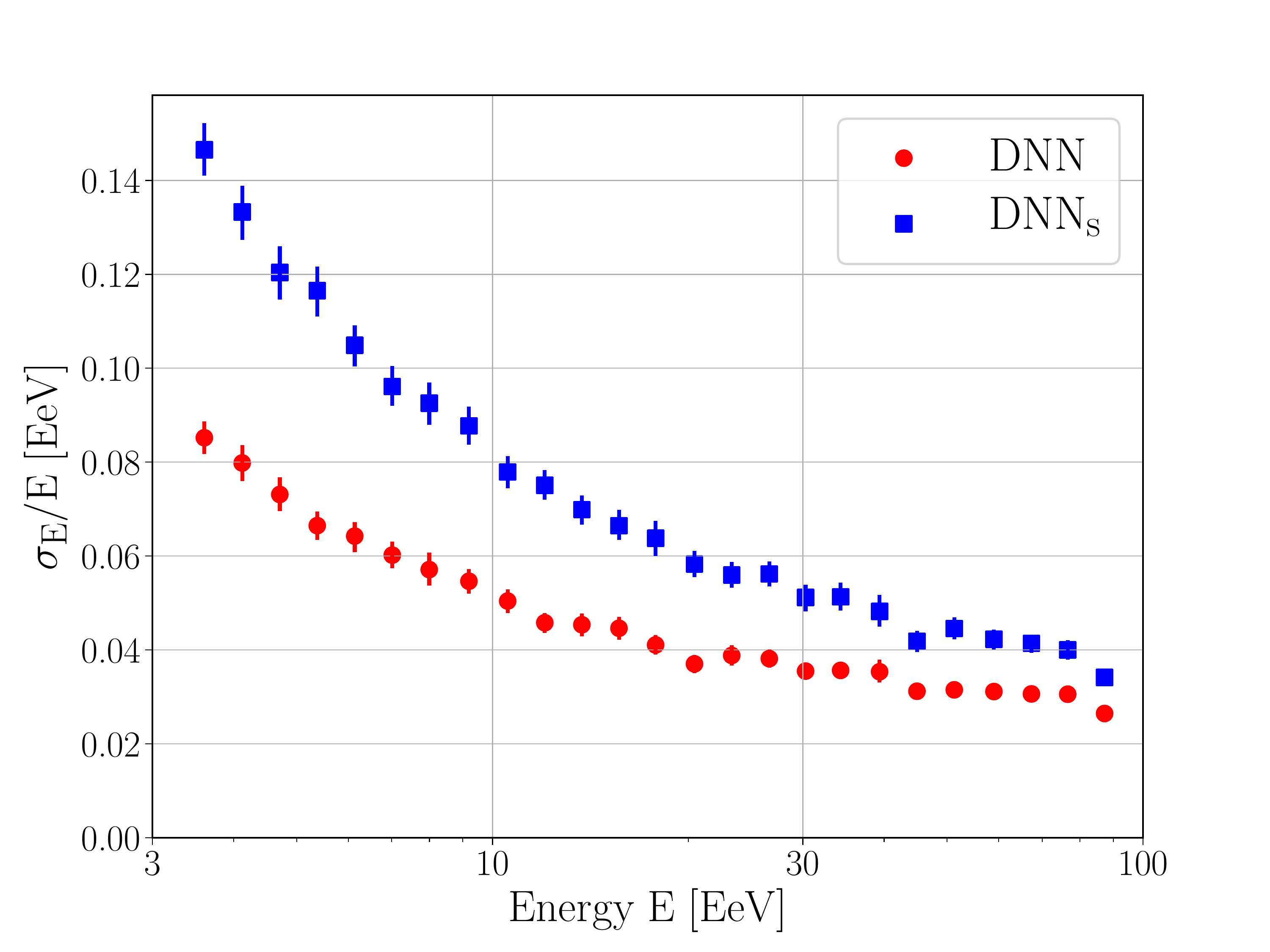}
\subcaption{}
\label{fig:energy_b}
\end{subfigure}
\caption{
(a) Difference between reconstructed and true cosmic ray energy for \SI{10}{EeV}. 
(b) Relative energy resolution as a function of cosmic ray energy. 
Symbols and curves refer to the network with information on total signal only (blue square symbols, dashed curve) and to the network with full information (red circular symbols, solid curve).
Error bars give the statistical uncertainty of the resolution.
}
\end{centering}
\end{figure*}

To investigate the performance of the network, initially we provide on input the same arrival times as for the plane fit.
The locations of the detectors are not provided and need to be inferred from the training data instead.

In Fig.~\ref{fig:angular_a}, the angular distance between the reconstructed and the true directions of the showers are shown by the dashed blue curve (DNN$_{\rm t}$).
Without explicit information on the physical conditions, e.g. locations of the detectors or velocity of shower particles, the network achieves the same angular resolution as the plane fit.

We also exploit information on the shower development that is contained in the time traces recorded by the detectors in addition to the arrival times of the first particles. 
In Fig.~\ref{fig:angular_a} we show the angular resolution of the combined information of arrival times and time traces (DNN, solid red curve). The angular resolution improves to \SI{1.2}{\degree}.

Note that the data set contains a substantial noise component such that the results of the plane fit are disturbed.
The network DNN is able to correct for these effects using additional information contained in the lateral signal distribution and the time traces.

In Fig.~\ref{fig:angular_b} we also show the energy dependence of the angular resolution. 
As expected, the angular resolution of the network DNN with the entire available input information provides a consistently better performance than the network DNN$_{\rm t}$ working only on the arrival times.
Compared to the plane fit, the network DNN performs equally well or better up to the highest energies.

\section{Cosmic ray energy}
Information on the cosmic ray energy is primarily contained in the lateral distribution of the shower particles and the total signals deposited in the detectors~\citep{Auger}.
We investigate energy reconstruction in two variants of the input data available to the main network.

Initially, we exclusively provide the two-dimensional distribution of the total signals (\ref{eq:Stot}) on input. In Fig. \ref{fig:energy_a} we show the reconstructed energy for cosmic rays with the mixed composition and with true energy \SI{10}{EeV} as the dashed blue curve (DNN$_{\rm s}$). 
The energy reconstruction exhibits a small bias and has a resolution of \SI{0.82}{EeV}, or $\sigma_E/E=8.2\%$, respectively.
Note that the absolute value of the energy resolution is dependent on the exact details of the shower and the detector simulation, and is used here only as a benchmark for further comparisons.

In a second step, on input we provide all other features in addition to the total signals, namely the arrival times (\ref{eq:t0}) and the $10$ features $f_j$ extracted from the signal amplitudes of the time traces. 
In Fig.~\ref{fig:energy_a} we show the energy resolution for the \SI{10}{EeV} cosmic rays by the solid red curve (DNN).
Using the full information, the energy resolution is reduced to $\sigma_E/E=5.1\%$.

\begin{figure*}[t!]
\captionsetup[subfigure]{aboveskip=-1pt,belowskip=-1pt}
\begin{centering}
\begin{subfigure}[b]{0.495\textwidth}
\includegraphics[trim={0 0 2cm 2.1cm},clip,width=\textwidth]{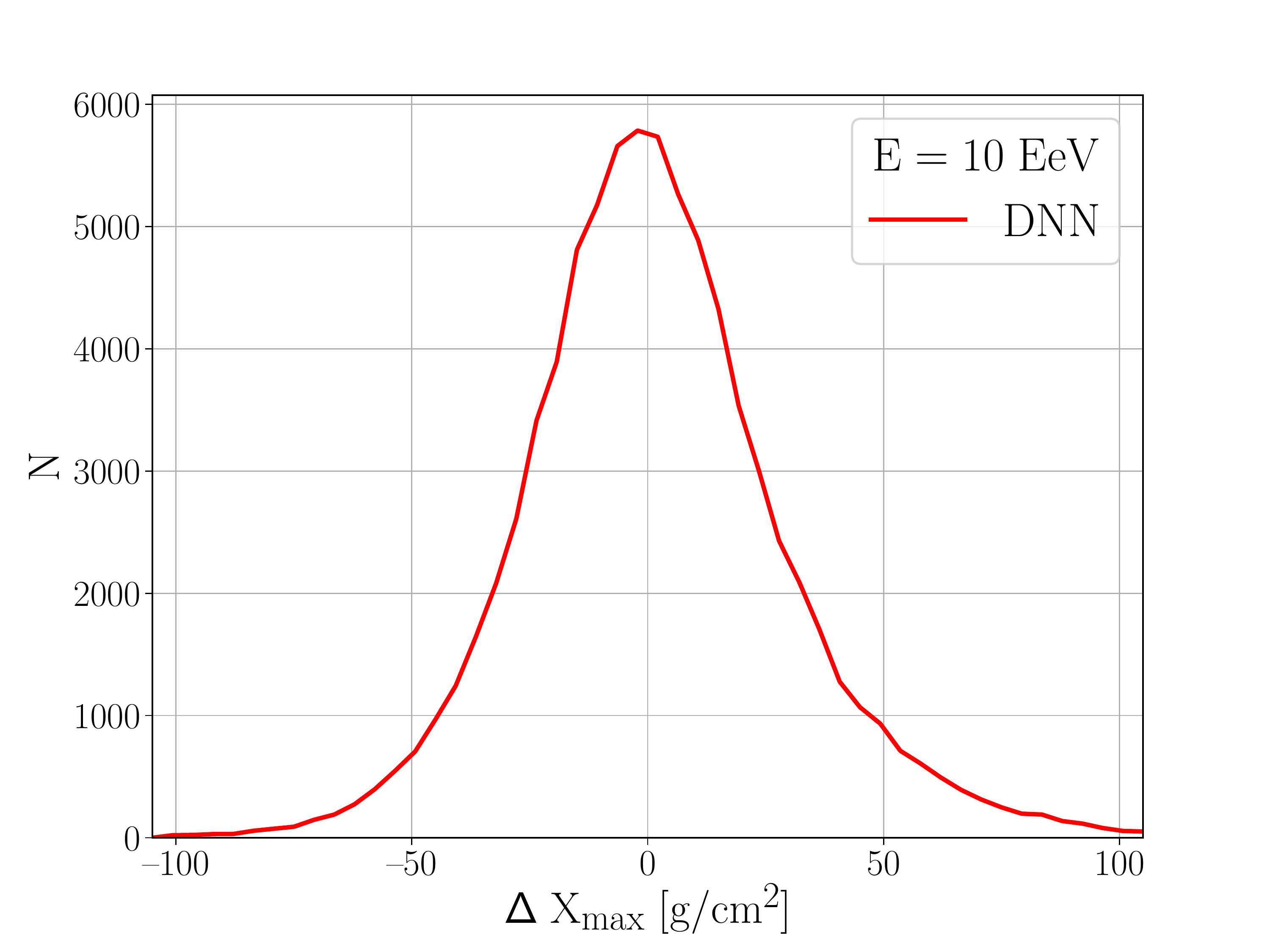}
\subcaption{}
\label{fig:Xmax_a}
\end{subfigure}
\hfill
\begin{subfigure}[b]{0.495\textwidth}
\includegraphics[trim={0 0 2cm 2.1cm},clip,width=\textwidth]{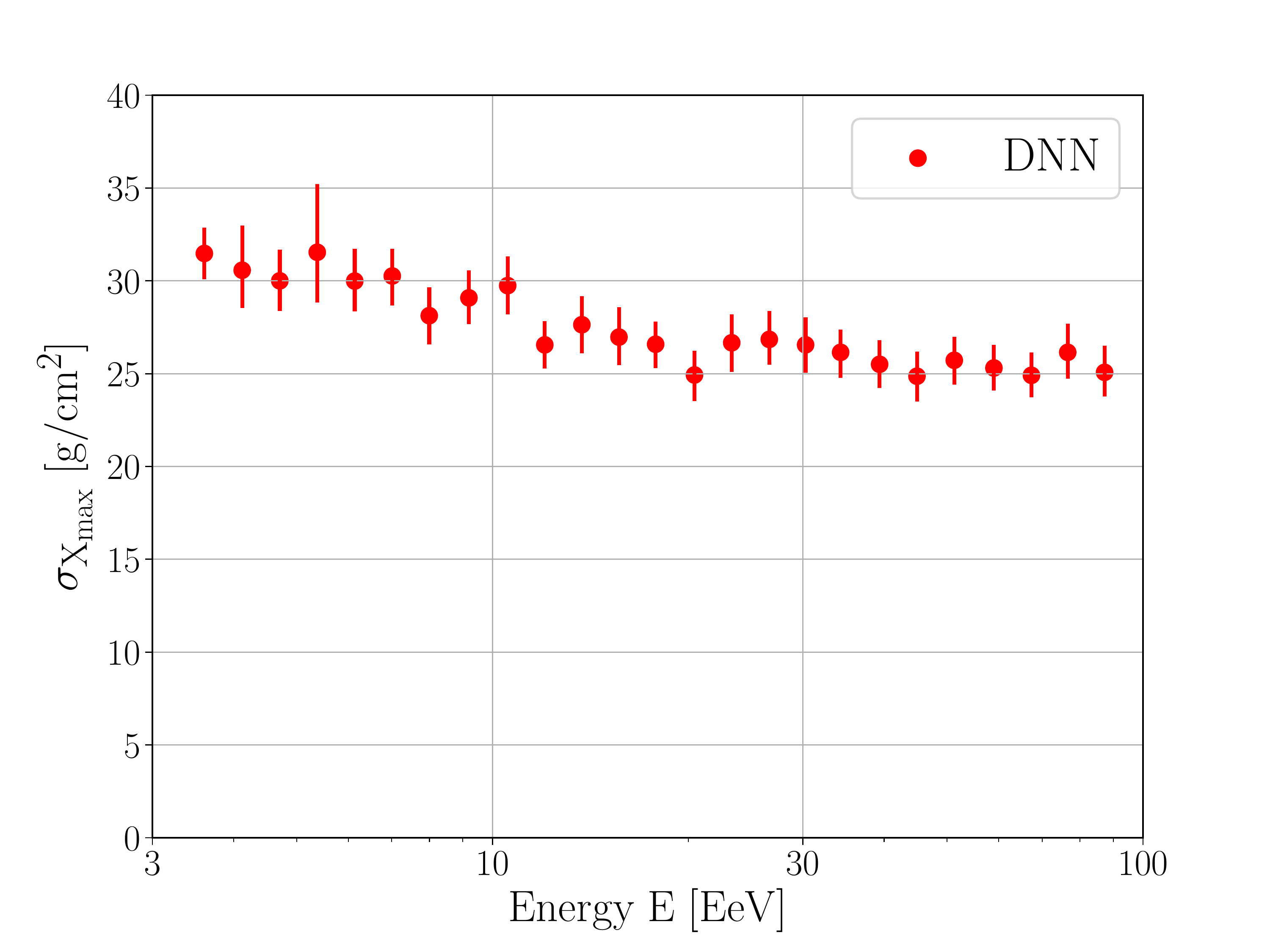}
\subcaption{}
\label{fig:Xmax_b}
\end{subfigure}
\end{centering}
\caption{
(a) Difference between reconstructed and true shower depths \Xmax for cosmic rays with \SI{10}{EeV} using the network with full information. 
(b) The \Xmax resolution as a function of the cosmic ray energy.
Error bars give the statistical uncertainty of the resolution.
}
\end{figure*}

In Fig.~\ref{fig:energy_b} we show the energy dependence of the energy resolution.
As we have parameterized with our simulation a calorimetric energy measurement, the relative energy resolution improves with increasing energy as the number of shower particles increases.

The resolution when using the full information (DNN, red circular symbols) is consistently better than the resolution when using the total signal distributions only (DNN$_{\rm s}$, blue square symbols).
This comparison demonstrates that the network DNN is able to take advantage of information contained in the time traces as well.

\section{Reconstruction of the shower depth}
Information on the cosmic ray composition is contained in the details of the shower properties, where a good estimator is the atmospheric depth of the maximum \Xmax of the shower development. 
Heavy cosmic nuclei typically interact high up in the atmosphere and produce a substantial number of muons.
In contrast, showers of light nuclei such as protons penetrate deeper into the atmosphere. They exhibit large fluctuations in \Xmax and produce less muons. 

Note that in our air shower simulation, the primary information on \Xmax is encoded in the time
traces as we approximate the shower front by a plane wave.
We again investigate air showers with the mixed composition of \ce{H}, \ce{He}, \ce{N}, \ce{Fe} with equal fractions.

In Fig.~\ref{fig:Xmax_a} we show the difference between the reconstructed and the true shower maxima \Xmax for cosmic rays with \SI{10}{EeV} energy. 
Here all information, the $10$ features $f_j$ plus the total signals and the arrival times are provided to the network (DNN). 
This demonstrates that the network is able to extract the relevant information from the data in order to reconstruct the shower maximum. 
Note also that the resolution of the \Xmax depends on the details of the air shower simulation.

In Fig.~\ref{fig:Xmax_b} we also show the energy dependence of the \Xmax resolution.
The shower footprint increases with increasing energy, allowing to better probe the different lateral distributions of the electromagnetic and muonic components, cf. Fig.~\ref{fig:trace}.
Additionally, random fluctuations and relative noise in the time traces decrease with increasing energy.
As a result, the reconstruction quality improves slightly with increasing cosmic ray energy.

\section{Conclusions}

The footprint of an extended air shower in an Earth-based observatory is like a very short video of the detector responses to traversing shower particles.
We developed a parametrized air shower and detector simulation to investigate reconstruction of shower properties from the detector measurements with convolutional network techniques.
To provide suitable input to our main network we casted the measured information on a regular two-dimensional (2D) grid of the detector stations.

The arrival times of the first shower particles at the detectors match on such a 2D grid.
This time information was sufficient for the network with dense 2D architecture to provide estimates of the shower directions with a resolution competitive to that of a classical fit to the particle shower front.
While prior knowledge was input on the exact detector locations and on the particle velocities for the plane fit, the network had no additional information beyond the arrival times.

For reconstructing the shower energy, the time-integrated signal amplitudes in the detectors can also be casted directly on the 2D grid.
To include also the time traces of the signal amplitudes themselves further steps are required. The traces contain information on the longitudinal shower development including the shower depth and the electromagnetic and muonic components of the shower.
Instead of directly using the amplitudes in the time bins specified by the hardware, we dedicate a part of the network to extract a small number of characteristic features describing the shape of the time traces.
Since the entire network is optimized as a whole, this network part is trained to extract those features which are most useful for the given reconstruction objective.
Every single feature was then casted on the 2D grid.

Using on input the above-mentioned arrival times, the time-integrated signal amplitudes, and all features extracted from the time traces, we again reconstructed the shower direction, energy and the shower depth with the dense 2D network. 
This set of input information yielded the best results. 
All shower properties were reconstructed with good precision and exhibited the expected dependency on the cosmic ray energy.
Without prior information on the physics of air showers, the network extracted all necessary information from the data provided for training purposes.

\section*{Acknowledgments}

This work is supported by the Ministry of Innovation, Science and Research of the State of North Rhine-Westphalia, and the Federal Ministry of Education and Research (BMBF).
We wish to thank Thorben Quast for his valuable comments on the manuscript.

\end{document}